\documentclass[twocolumn, secnumarabic, nobibnotes, aps, prd]{revtex4-2}

\usepackage{amsmath}
\usepackage{amsfonts}
\usepackage{amssymb}
\usepackage{graphicx}
\usepackage{soul}

\usepackage[utf8]{inputenc}
\begin{document}

\title{Staircase penetration of magnetic flux into a superconducting flat ring}
\author{Leonid Burlachkov}
\email{leonid@biu.ac.il}
\author{Nikita Fuzailov}
\email{fuzailovnikita@gmail.com}
\affiliation{Department of Physics, Bar-Ilan University, Ramat-Gan 5290002, Israel}

\begin{abstract}
We analyzed the distribution of the Meissner shielding currents in a flat superconducting ring and quantitatively described the penetration of magnetic avalanches (dendrites) inside it. Using a recurrent procedure, the external field $H_{ext}$, in which a perforating (edge to edge crossing) avalanche appears, is calculated. A staircase dependence of the mean field trapped inside a ring, $\left\langle H\right\rangle$, vs. $H_{ext}$, is comprehensively described. A staircase slope appears to be a universal function of a ring shape (inner to outer diameter ratio). The heat released by a penetrating dendrite grows linearly with each next perforation. Flux pinning, if present, modifies a staircase dependence and makes steps smaller, but does not change the staircase slope.  Our theoretical results are in a good accordance with the experimental data. 
\end{abstract}

\maketitle

\section{Introduction}

Magnetic flux penetration into flat superconducting samples, whose thickness is much less than their in-plane dimensions, is quite fascinating and even curious. A weak external magnetic field $H_{ext}<H_{c1}$, where $H_{c1}$ is the first critical field, applied perpendicularly to a sample, is repealed from its bulk (the Meissner effect) but accumulated at the periphery since the demagnetizing factor $ N$ is close to $1$ for flat samples. Correspondingly, the magnetic field near the outer edges is much greater than the external field: $H_{edge}\simeq H_{ext}/(1-N)\gg H_{ext}$. As $H_{ext}$ ramps up, magnetic vortices break inside through a surface barrier, which is of a geometrical type and inevitably present in flat samples \cite{Zeldov}, though the Bean-Livingston barrier \cite{BeanLivingston,BurlachkovBL} can be also essential. Penetration takes place at distinct points of the outer edge where a barrier is weakened by any reason: geometric defects, composition imperfections, and so on. As a result, magnetic flux penetrates into a sample in a captivation form of dendrites, lightnings or "fingers", which can be visualized by a magneto-optical technique \cite{Alvarez,Polturak,Baruch-El1,Qureishy,Baruch-El2,Brisbois,Baziljevich,Vestgarden,Aranson}. It is worth mentioning that such an avalanche-like flux penetration has a lot in common with formation of discharge channels at dielectric breakdown \cite{Niemeyer}, and in general can be described by the theory of self-organized criticality \cite{Bak,Wijngaarden}.

Though most of experiments with dendrite and other avalanche-like flux penetration were performed using flat samples of rectangular shape or disks, a special attention was triggered to rings $b\leq r\leq a$ \cite{Shvartzberg,Olsen1,Olsen2,Pannetier,Vodolaz,Jiang}, where $b$ and $a$ are the inner and outer radii, respectively (see Fig.\,1). Flux dendrites, or avalanches (sometimes they almost do not bend and/or branch and can be called "fingers") penetrate into a ring from its outer edge and either reach the inner edge or end up somewhere within the ring body. If a finger perforates (crosses) the entire ring, the effect is equivalent to building up a magnetic bridge between the outer and the inner edges. When such a perforation happens for the first time, the full Meissner shielding of the inner part $r<b$ gets broken and magnetic flux penetrates inside in an form of a vortex avalanche. At the same time, the Meissner state $B=0$ inside the ring body ($b\leq r\leq a$) is, of course, preserved except for the path of the bridge. After the avalanche is over, the magnetic bridge disappears, and $H_{ext}$ should be increased further until the next perforation occurs. Unlike the Meissner state in a superconducting hollow cylinder, where $H=0$ everywhere inside the cavity, the magnetic field in the inner part of a flat ring is never uniform. Thus, it is more convenient to describe the field penetration in terms of the total magnetic flux trapped inside a ring (at $r<b$)%
\begin{equation}
\Phi=2\pi \int\limits_{0}^{b}B(r)\,r\,dr  \label{eq-flux-hole}
\end{equation}
and the mean trapped field $\left\langle H \right\rangle =\Phi/\pi b^{2}$. Of course, at low $H_{ext}$, before the first magnetic avalanche penetrates inside the ring from edge to edge, both $\Phi$ and $\left\langle H \right\rangle $ are zero. As $H_{ext}$ grows up, each next perforating avalanche leads to a step-like increase of $\Phi$, which results in a staircase-like dependence of $\Phi$ on $H_{ext}$ \cite{Olsen1,Shvartzberg,Jiang}. It is worth emphasizing that appearance of a perforating avalanche does not imply establishing a condition $\left\langle H \right\rangle =H_{ext}$. This would be the case for a long hollow cylinder with a slit, but in a flat ring we always get $\left\langle H \right\rangle >H_{ext}$ due to a "flux-focusing" effect \cite{Brojeny,BrandtClem}.

The goal of our paper is to construct a comprehensive theory of a discrete (staircase-like) flux penetration into a type-II superconducting ring, which should be valid at any ratio $b/a$, including thick ($b\ll a)$ and thin ($a-b\ll a$) rings. In the previous studies \cite{Olsen1,Shvartzberg,Jiang}, an avalanche-like penetration was described in terms of the Bean \cite{Bean} or Bean-like (Kim \cite{Kim}) model of a critical state. This implies that the field distribution along an avalanche, $B(r)$, is determined by the relation $dB(r)/dr \propto j_{c}(B)$, where $j_{c}$ is the critical current density. To our mind, though the Bean model proves to successfully describe a flat-front penetration of magnetic flux into disks and rings \cite{Brandt1997}, a possibility to incorporate a spike-like lightening into a critical-state description looks questionable. Alternatively, our approach is not based on any critical state or other specific model. We study an avalanche-like flux penetration into a superconducting ring from first principles by considering an interplay between the magnetic self-energy of a dendrite (finger) and its Lorentz energy. The latter emerges from the interaction between the penetrating magnetic flux and the Meissner shielding currents. In general, our approach is analogous to a classic calculation of the first critical field $H_{c1}$ in a bulk superconductor, but generalized for the case of a linear structure of penetrating flux (dendrite) and a flat ring with $a\gg s$, where $s$ is its thickness. Such an approach enables calculation of magnetic field penetration into a "clean", pinningless ring, and accounting for pinning as well. In both cases (with or without pinning) we get a staircase-like dependence of $\Phi$ on $H_{ext}$. In terms of the ring thickness, we confine ourselves by the case $s>\lambda$, where $\lambda$  is the in-plane London penetration depth. The opposite case $s<\lambda$ (Pearl vortices) requires separate analysis, though the results obtained in Refs.\,\cite{Brojeny,BrandtClem} suggest that both cases are similar if one substitutes the Pearl effective penetration depth $\Lambda =\lambda ^{2}/s$ for $\lambda$. Anyway, in all the experimental works we refer to in this paper, the condition $s>\lambda $ holds.

It is worth emphasizing that a superconducting ring, especially a set of equally prepared rings with different radii $a$ and $b$, provides a unique opportunity to study magnetic properties of type-II superconductors. Unlike disks or strips, avalanches penetrating into rings result in step-like changing of macroscopic and easily measurable properties such as the mean field trapped inside the ring, magnetic moment, torque, etc. A superconducting ring allows us to digitize macroscopic quantum effects, thus a comprehensive study of magnetic perforation into rings is so important and challenging.

\section{Energy of a propagating dendrite (finger) without pinning}

Consider a thin pinningless superconducting ring $b<r<a$ of thickness $s\ll a$ embedded into a perpendicular magnetic field $H_{ext}$, as shown in Fig.\,1. 
\begin{figure}[t!]
    \centering
    \includegraphics[width=0.48\textwidth]{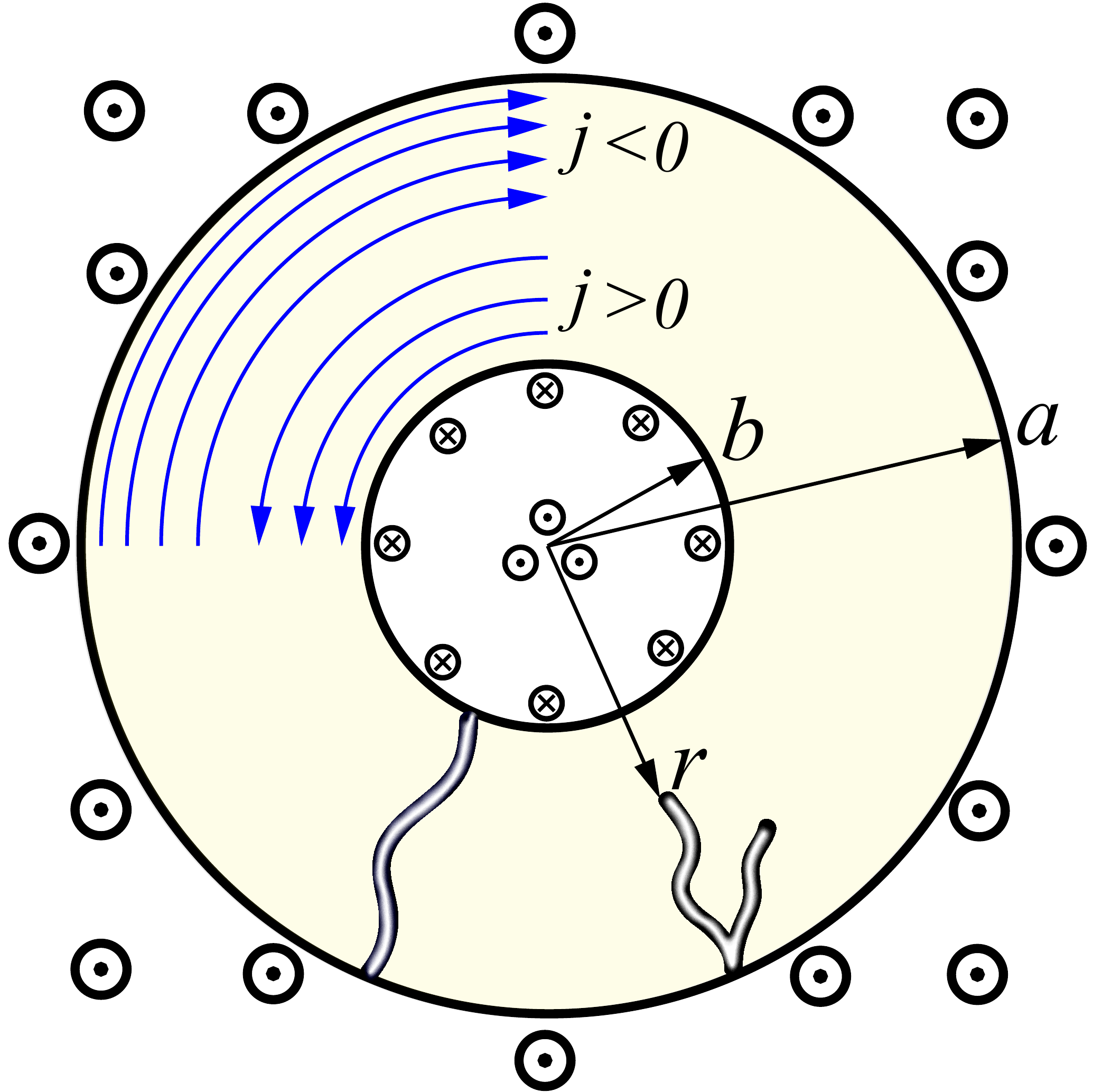}
    \caption{Flux avalanche (dendrite) penetration into a superconducting ring. The counterclockwise Meissner currents are positive since they create and upward-directed magnetic field inside the ring (at $r<b$), the clockwise currents are negative. Circles denote magnetic field directed inside ($\otimes$) and outside ($\odot$) the drawing plane.}
\end{figure}
The total current in the ring is $I=\int_{b}^{a}\,j(r)\,dr$, where $j(r)$ is a linear current density measured in $\mathrm{A/cm}$. Note that a superconducting current flows in the $\lambda$ surface layers, so most of the ring's bulk is free of current since $s\gg \lambda $. Thus, it does not make sense to speak about a true current density $j/s$. The counterclockwise currents, which produce an upward magnetic field in the central part $r<b$, see Fig.\,1, are considered positive. Correspondingly,  the clockwise currents are negative. The field at the ring center $r=0$ is
\begin{equation}
H_{center}=H_{ext}+\frac{2\pi }{c}\int_{b}^{a}\,j(r)\,\frac{dr}{r},
\label{eq-Hcenter}
\end{equation}%
whereas $H(r)$ at $r \neq 0$ can be calculated numerically. Of course, the ring body is always in the Meissner state: $H=0$ at $b<r<a$.

In order to set up a numerical problem, we introduce the dimensionless variables: $\beta =b/a$, $\zeta =r/a$, (so $\beta \leq \zeta \leq 1$), $h=H/H_{0}$ and $\tilde{j}=j/j_{0}$. Here
\begin{equation}
H_{0}=\frac{\phi _{0}s\ln \kappa }{8\pi a\lambda ^{2}}=\frac{H_{c1}s}{2a},\quad j_{0}=\frac{H_{0}c}{2\pi },  \label{eq-H0j0}
\end{equation}%
$c$ is the speed of light, $\phi_{0}=\pi \hbar c/e$ is the quantum flux, $\kappa =\lambda /\xi $ is the Ginzburg-Landau parameter and $\xi $ is the superconducting coherence length. In our further numerical analysis we use a straightforward matrix-inversion method \cite{BrandtClem} in order to calculate $h(\zeta)$, $\tilde{j}(\zeta)$ and all other values. A propagating dendrite is approximated as a linear finger of width $w$ which emerges from the outer edge $r=a$ and grows toward the inner edge $r=b$. We discuss the effect of possible branching and bending in Section VII, which is devoted to comparison of our results to experimental data. If $\rho$ is the concentration of vortices in such a finger, and its innermost part is located at distance $r$ from the ring center, thus the finger length is $a-r$, see Fig.\,1, and its self-energy can be written as%
\begin{equation}
E_{self}=qs \rho w\varepsilon_{0}(a-r)=qs \rho w\varepsilon _{0}a(1-\zeta ).
\label{eq-Eself}
\end{equation}%
Here $\varepsilon _{0}=(\phi _{0}/4\pi \lambda )^{2}\ln \kappa =\phi_{0}H_{c1}/4\pi $ is vortex energy per unit length. The coefficient $q(\rho)\geq 1$ accounts for interaction (repulsion) between vortices and for some bending and branching of a finger. If a finger is straight with no branching, then $q\varepsilon_{0}$ is just its self-energy energy per one vortex. For a straight finger with a low lattice density, $\rho <1/\lambda^{2}$, which implies that the mean distance between vortices is greater than $\lambda$, we apparently have $q=1$. However, as shown below is Section VII, $q\gg 1$, which proves that the vortex structure in a finger is rather dense. This fact looks quite natural, since the mean intervortex distance in a fast-propagating avalanche with strong interaction between vortices is expected to be less than $\lambda$.

Another contribution to a finger energy comes from the Lorentz force equal to $f_{L}=\phi _{0}j/c$ per one vortex. Note that a positive (counterclockwise) current pushes vortices toward the outer edge $r=a$, whereas a negative (clockwise) current forces them toward the ring center. Both positive and negative $j$ can coexist in a ring at the same time as shown in Fig.\,1. The Lorentz energy of a single vortex located at a distance $r$ from the center is%
\begin{equation}
u_{L}(r)=\frac{\phi _{0}}{c}\int_{r}^{a}j(r_{1})\,dr_{1}=\frac{\phi
_{0}aH_{0}}{2\pi }\int_{\zeta }^{1}\tilde{j}(\zeta _{1})\,d\zeta _{1}\ .
\label{eq-Esingle}
\end{equation}%
Thus the total Lorentz energy of a finger of length $a-r$ is
\begin{equation}
\begin{split}
E_{L}(r) = \rho w\int_{r}^{a}u_{L}(r_{1})\,dr_{1}=  \\
\frac{\phi _{0}a^{2} \rho wH_{0}}{2\pi }\int_{\zeta }^{1}d\zeta _{1}\int_{\zeta
_{1}}^{1}d\zeta _{2}\,\,\tilde{j}(\zeta _{2})\,.
\end{split}
\label{eq-EL}
\end{equation}%
Using Eqs.\,(\ref{eq-Eself}) and (\ref{eq-EL}), we express the total energy of a finger as a function of $\zeta =r/a$ as
\begin{equation}
E(\zeta )=E_{self}(\zeta)+E_{L}(\zeta)=E_0 \tilde{E}(\zeta ),
\label{eq-E}
\end{equation}%
where%
\begin{equation}
\tilde{E}(\zeta )=q(1-\zeta )+\int_{\zeta }^{1}d\zeta _{1}\int_{\zeta
_{1}}^{1}d\zeta _{2}\,\,\tilde{j}(\zeta _{2})  \label{eq-Etilde}
\end{equation}%
is the dimensionless energy of a finger (dendrite) of the length $1-\zeta$ and $E_0=\phi _{0}a \rho wsH_{c1}/4\pi$. Note that our calibration of $H_{0}$, see Eq.\,(\ref{eq-H0j0}), results in the most compact expression for $E$.

Apparently, in the absence of pinning forces, a penetrating finger appears if $\tilde{E}(\zeta )$ is a monotonously increasing function at $\beta \leq \zeta \leq 1$, which means
\begin{equation}
\frac{d\tilde{E}}{d\zeta }=-q-\int_{\zeta }^{1}d\zeta _{1}\,\,\tilde{j}%
(\zeta _{1})>0,  \label{eq-dEdzeta}
\end{equation}%
see Fig.\,2. 
\begin{figure}[t!]
    \centering
    \includegraphics[width=0.48\textwidth]{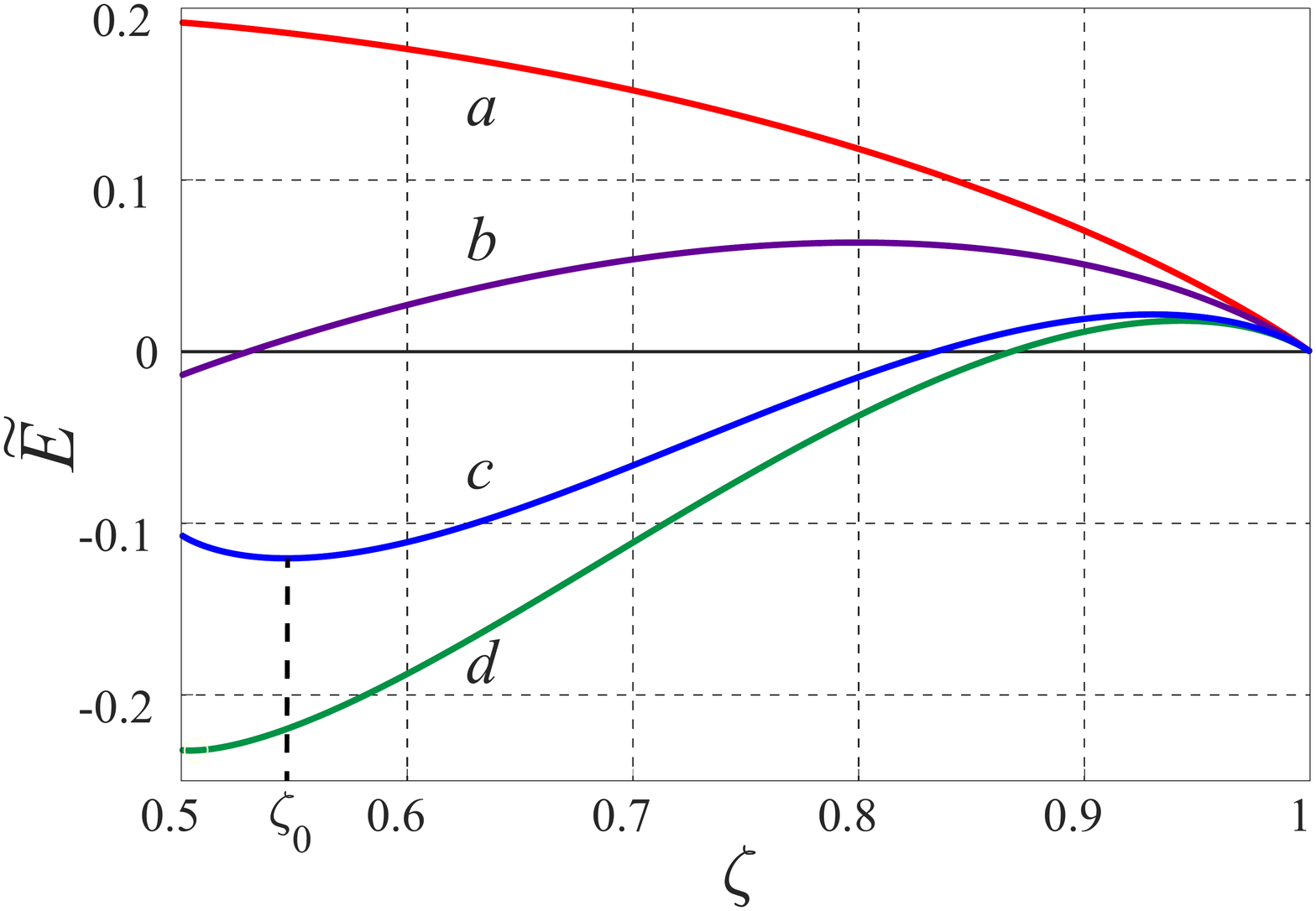}
    \caption{Energy $\tilde{E}(\zeta)$ of a penetrating dendrite (except the first one) for $\beta=0.5$, see Fig.\,1 and Eq.\,(\ref{eq-Etilde}). (a) $\tilde{E}>0$ in the whole ring width, no penetration; (b) $\tilde{E}(\beta)<0$, but a huge potential barrier (where $\tilde{E}>0$) is present, no penetration; (c) potential barrier is surmountable, but an avalanche propagates till $\zeta_0$ where $\tilde{E}(\zeta)$ has a minimum; (d) The perforation condition, see Eq.\,(\ref{eq-perfor}), is fulfilled, $d\tilde{E}/d\zeta \mid _{\zeta =\beta}=0$, and an avalanche perforates the ring from the outer edge to the inner one. }
\end{figure}
However, this condition is never satisfied at the outer edge $\zeta =1$ of a ring. As was mentioned in numerous studies starting from the pioneer work of Larkin and Ovchinnikov \cite{Larkin}, the Meissner shielding current diverges as $(1-\zeta )^{-1/2}$ near the edge of a flat superconductor, no matter if it is a disk, ring or stripe. Correspondingly, the integral in Eq.\,(\ref{eq-dEdzeta}) turns to zero as $(1-\zeta )^{1/2}$ when $\zeta \rightarrow 1$, and $d\tilde{E}/d\zeta \mid _{\zeta =1}=-q<0$. This implies an inevitable existence of a potential barrier for a finger emergence at an outer edge, since its (positive) self energy $E_{self}$ grows linearly with its length $1-\zeta $, see Eq.\,(\ref{eq-Eself}), whereas the (negative) Lorentz energy, described by Eq.\,(\ref{eq-EL}), is proportional to $(1-\zeta )^{3/2}$. As a result, any appearance of a propagating finger, whether it is perforating (reaching the inner edge of a ring) or not, is an activation process. Almost in all experimental visualizations of magnetic flux penetration into flat superconducting samples \cite{Alvarez,Polturak,Shvartzberg,Baruch-El1,Qureishy,Baruch-El2,Brisbois,Baziljevich,Vestgarden} just a few fingers (dendrites, lightnings) are observed in a macroscopic sample at the same time. This confirms the fact that a vortex avalanche has to surmount a macroscopic potential barrier at an early stage of its formation. Therefore, such avalanches originate from distinct points at the outer edge where geometric or material imperfections are present, and the surface barrier is locally suppressed.

Assume that at certain $h_{ext}=h_1$ the surface barrier becomes locally surmountable, thus an avalanche emerges at the outer edge $\zeta =1$ and propagates inside. As we show below in Section IV, in the absence of pinning the first penetrating finger is already a perforating one, i.e., it should reach the inner edge $\zeta =\beta $, see Fig.\,1. A partial penetration, shown in the same Fig.\,1, is possible at further stages. A perforating finger serves as a magnetic "bridge" connecting the inner and the outer edges of the ring. Vortices start to move along this bridge, magnetic flux in the central hole $\zeta<\beta$ increases until the bridge gets locked. It happens when the work of superconducting currents $j$ on a vortex which travels throughout the whole finger from $r=a$ to $r=b$ turns negative, or at $\tilde{I}=\int_{\beta }^{1}\,\tilde{j}(\zeta )\,d\zeta =0$. As soon as the "zero total current" state is established, flux penetration inside the ring along the first perforating finger stops, and $h_{ext}$ needs to be increased further until the next perforation happens.

It is worth emphasizing that we do not consider a vortex avalanche to destroy superconductivity inside its path. We assume that the shielding currents $j$ flow across the finger, whether it is perforating or not, and produce a Lorentz force on it as described by Eqs.\,(\ref{eq-Esingle}) and (\ref{eq-EL}). In terms of a current flow, the ring remains superconducting as a whole, $j$ is circular and crosses the avalanche path. There could be another glance on a dendrite penetration: one might consider the thermal instability, which accompanies such a vortex avalanche, to be strong enough in order to destroy superconductivity and create a magnetic slit along its path. Then a perforating finger just "cuts" the ring and produces an edge to edge slit which interrupts $j$. However, within this description the result will be obviously the same: flux penetration inside the ring stops at $\tilde I = 0$, since the total current in a slit ring is zero. However, we think that our approach, where avalanches do not destroy superconductivity, is more appropriate to describe the available experiments with superconducting rings \cite{Olsen1,Shvartzberg,Jiang}. The appearance of an almost straight avalanche with little branching, which propagates toward the ring center, requires a significant Lorentz force which pushes vortices inside. This implies that $j$ flows across the finger path. Alternatively, the case where dendrites are branching a lot and propagate in various directions other than toward the ring center, points to a "burnt out" superconductivity along a dendrite path.

"Flux pumping" inside a ring by a perforating finger (dendrite) completely changes the structure of currents $j(\zeta)$. The flux $\Phi$, penetrated into the inner area $\zeta < \beta$ of a ring, produces a positive (counterclockwise) current $j$ flowing around the inner edge, see Fig.\,1. This results in repelling vortices from the inner part of a ring and "locking" the ring for further perforation. At the same time, a partial (non edge-to-edge crossing) penetration of fingers is still possible up to $\zeta=\zeta_0$, see the blue curve in Fig.\,2. As $h_{ext}$ is increased further, negative (clockwise) $j$, which pushes the magnetic flux inside the ring, grows up till the condition described by Eq.\,(\ref{eq-dEdzeta}) is fulfilled again. This happens, if we leave aside the problem with the surface barrier in the area $\zeta \approx 1$, at 
\begin{equation}
\frac{dE}{d\zeta}\,\biggr|_{\zeta =\beta}=0,
\label{eq-dE0}
\end{equation}
see the green curve in Fig.\,2.
Using Eqs.\,(\ref{eq-dEdzeta}) and (\ref{eq-dE0}), we get the perforation condition:
\begin{equation}
\tilde{I}_{perf}=\int_{\beta }^{1}\,\,\tilde{j}(\zeta )\,d\zeta =-q.
\label{eq-perfor}
\end{equation}%
As Eq.\,(\ref{eq-perfor}) is satisfied at $h=h_{2}$, the next perforating finger appears. A magnetic bridge connects the outer and inner edges of the ring, additional portion of magnetic flux penetrates inside, $\tilde{I}$ turns to zero again, and $h_{ext}$ should be increased further until the penetration condition described Eq.\,(\ref{eq-perfor}) is fulfilled again at $h=h_{3}$. Then the next perforating avalanche takes place, and so on. Eventually, we come to a staircase-like flux penetration which is observed experimentally \cite{Olsen1,Shvartzberg,Jiang} and illustrated in Fig.\,3. Quantitatively such a staircase will be described in the next Section.
\begin{figure}[t!]
    \centering
    \includegraphics[width=0.48\textwidth]{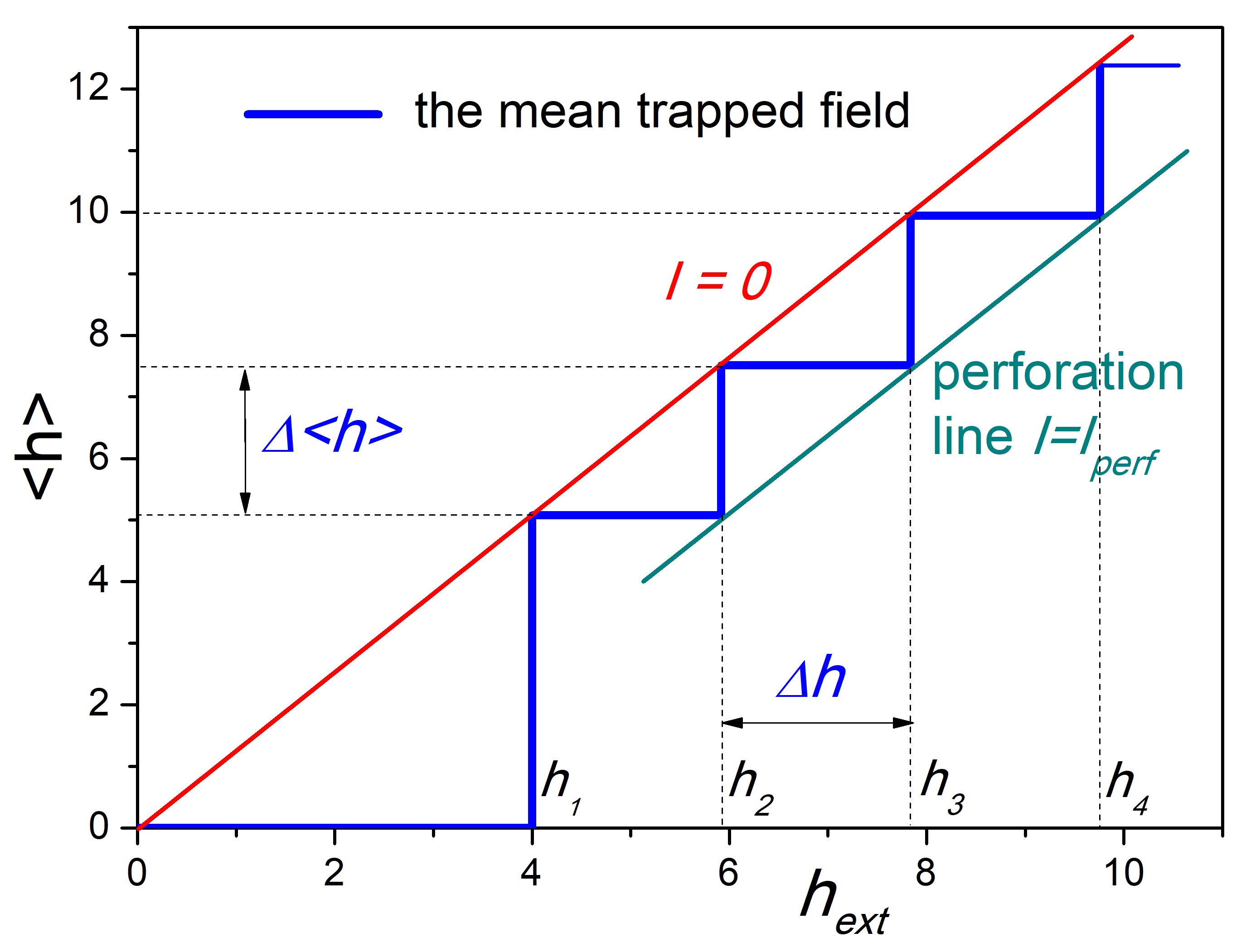}
    \caption{The mean field $\left\langle h\right\rangle $ trapped in a ring as a function of $h_{ext}$ at $\beta=0.75$ in a clean sample (no pinning). The first step (see Section IV) is chosen arbitrarily greater than the next identical ones, whose width and height, as well as the slope of $I=0$ line, are determined by Eq.\,(\ref{eq-staircase}). The axes scales correspond to $q=1$, so all the values (except the slope) should be multiplied by $q$. }
\end{figure}

\section{Recurrent calculation of a staircase flux penetration - no pinning}

In the previous section we explained qualitatively how a staircase dependence of $\left\langle h \right\rangle $ on $h_{ext}$ appears. Now we turn to its quantitative description. Let us define two currents: $\tilde{j}_{s}(\zeta )$ and $\tilde{j}_{p}(\zeta )$. The first one, $\tilde{j}_{s}$, creates a field $h_{s}=-1$ within the ring body (at $\beta \leq \zeta \leq 1$), whereas the total flux which it produces in the inner circle (at $\zeta <\beta $) is zero: $\tilde{\Phi}_{s}=2\pi \int_{0}^{\beta}h_{s}\,(\zeta )\,\zeta \,d\zeta =0$, compare to Eq.\,(\ref{eq-flux-hole}). The second current, $\tilde{j}_{p}$, produces a zero field $h_{p}=0$ inside the ring body and a flux $\tilde{\Phi}_{p}=2\pi \int_{0}^{\beta }h_{p}\,(\zeta )\,\zeta \,d\zeta =\pi \beta ^{2}$ in the inner circle. Note that $\tilde{\Phi}_{p}$ is just equal to the dimensionless area $\pi \beta ^{2}$ of the inner circle. The real flux defined by Eq.\,(\ref{eq-flux-hole}) is $\Phi =H_{0}\pi b^2\tilde{\Phi}$. At $h_{ext}<h_1$, where no vortex has yet penetrated inside the ring (full Meissner shielding), we have $\tilde{j}=\tilde{j}_s-\tilde{j}_p$, as shown in Fig.\,4. Obviously, only such a combination of $j_{s}$ and $j_{p}$ simultaneously satisfies two conditions: (1) $h=0$ inside the ring body; (2) $\tilde{\Phi}=0$. Note that $h(\zeta)$ in any case is a sum of $h_{ext}$ and the field generated by $\tilde{j}$, the same applies to $\tilde{\Phi}$. 
\begin{figure}[t!]
    \centering
    \includegraphics[width=0.48\textwidth]{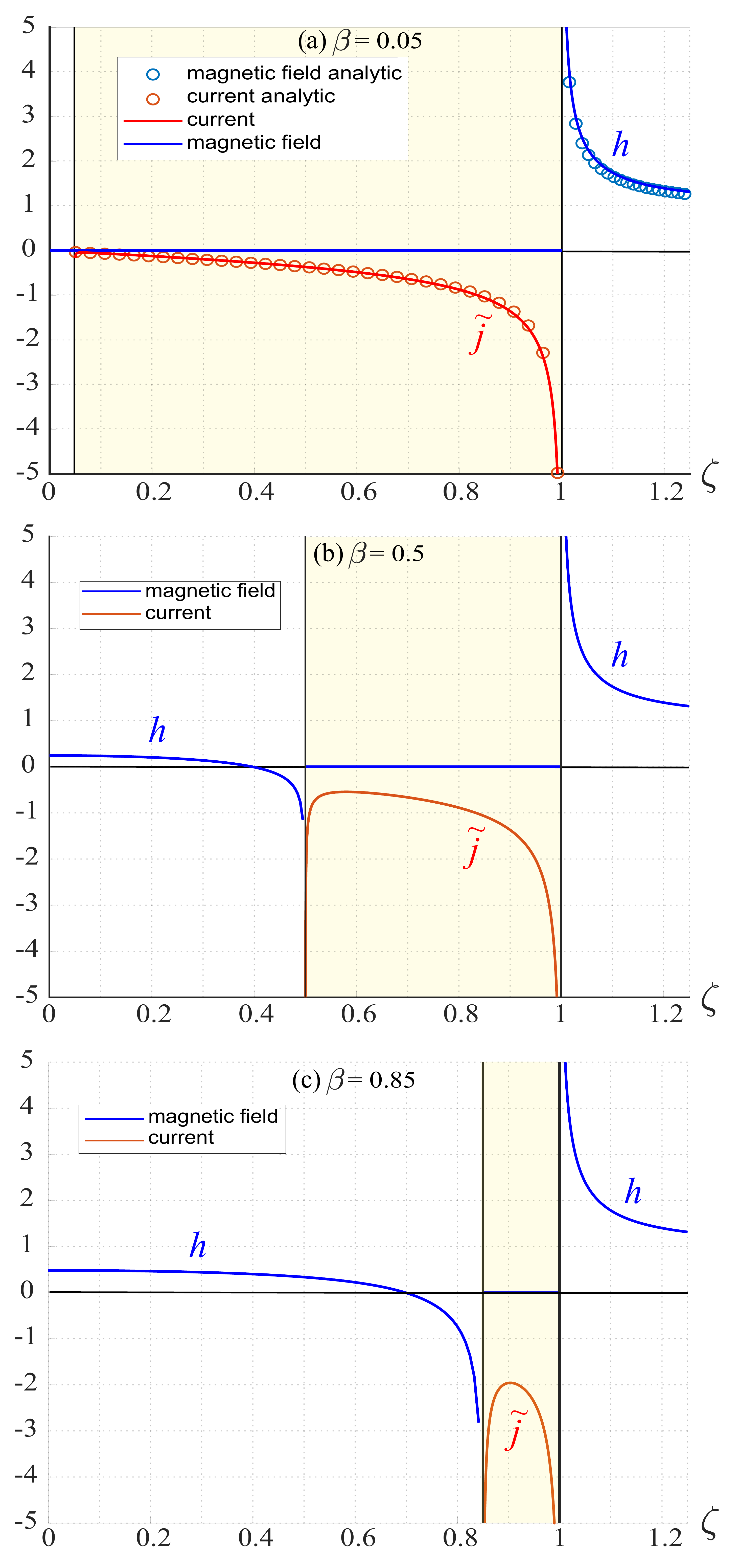}
    \caption{Numerical calculation of $h(\zeta)/h_{ext}$ and $\tilde{j}=\tilde{j}_s-\tilde{j}_p$ for a full shielding ($h_{ext}<h_1$, $\left\langle h\right\rangle=0$) at various ring widths $1-\beta$. Both $h$ and $\tilde{j}$ are dimensionless, as well as the vertical axis. For a very thick ring $\beta=0.05$ our numerical results are compared with the analytic expressions obtained for a disk \cite{Mikheenko}, see Eq.\,(\ref{eq-j-beta0}), and shown as circles. The field $h(\zeta)$ inside a ring (at $\zeta<\beta$) changes sign, and $\left\langle h\right\rangle =0$. The current $\tilde{j}(\zeta)$ is negative (clockwise).}
\end{figure}
At $\beta =0$ (superconducting disk) obviously we get $\tilde{j}_{p}(\zeta )=0$, and for $\tilde{j}_{s}(\zeta )$ there exists an analytic solution \cite{Mikheenko}:
\begin{equation}
\tilde{j}_{s}(\zeta )=-\,\frac{2}{\pi }\frac{\zeta }{\sqrt{1-\zeta ^{2}}},\qquad \beta =0,  \label{eq-j-beta0}
\end{equation}%
shown by circles in Fig.\,4(a). Note that Eq.\,(\ref{eq-j-beta0}) differs from the analogous expression presented in Ref.\,\cite{Mikheenko} by a factor of 2 due to a different calibration of $H_{0}$ chosen in our paper, see Eq.\,(\ref{eq-H0j0}). 

We denote the corresponding total currents as $\tilde{I}_{s}(\beta )=\int_{\beta }^{1}\,\,\tilde{j}_{s}(\zeta )\,d\zeta$ and $\tilde{I}_{p}(\beta )=\int_{\beta }^{1}\,\,\tilde{j}_{p}(\zeta )\,d\zeta$. These currents are numerically calculated for the range $0.1<\beta <0.95$
\begin{figure}[t!]
    \centering
    \includegraphics[width=0.48\textwidth]{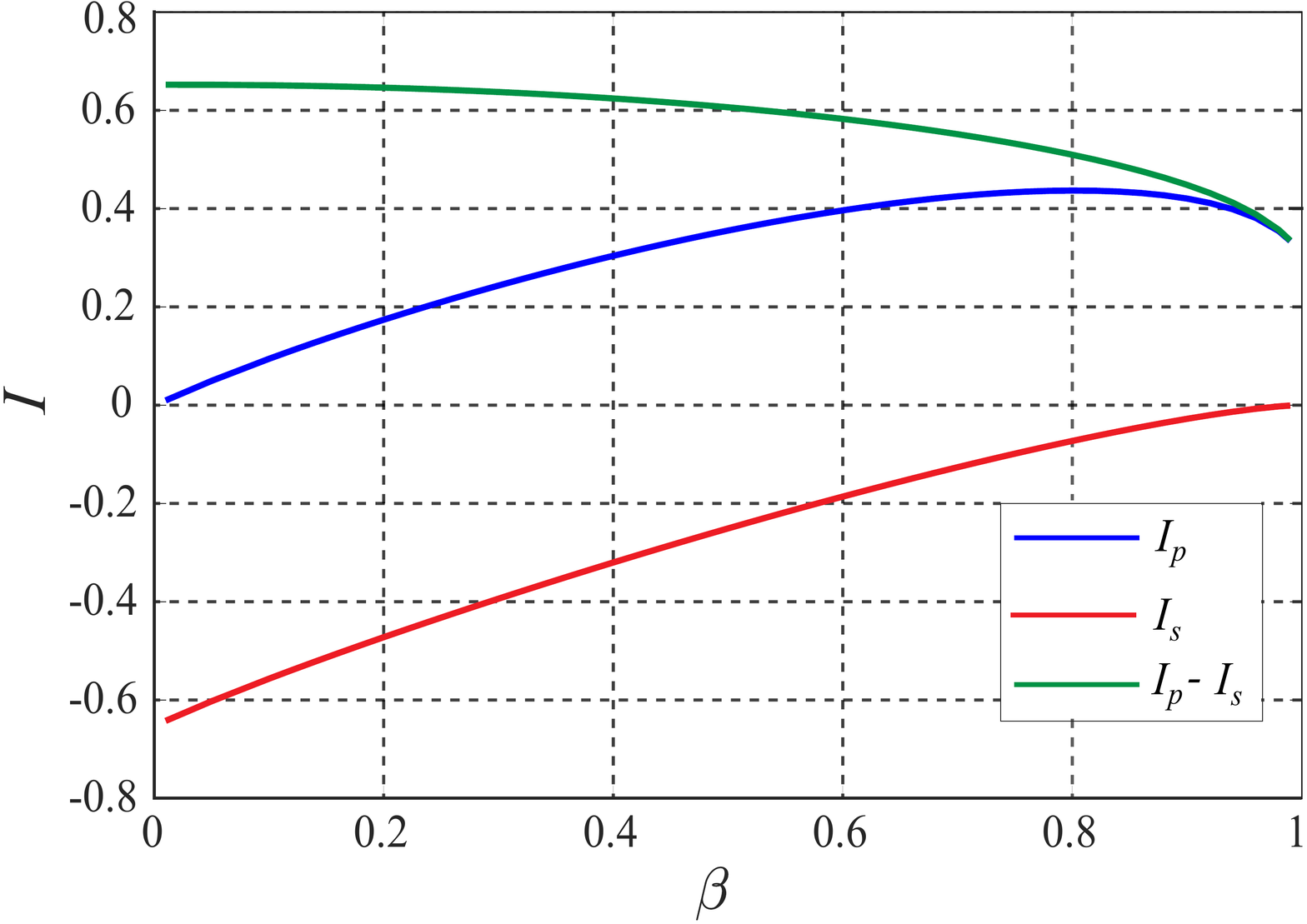}
    \caption{The currents $\tilde{I}_{s}$, $\tilde{I}_{p}$ and their difference $\tilde{I}_{p}-\tilde{I}_{s}$ as functions of $\beta$. These currents determine the staircase structure: its slope and the width and height of steps, see Eq.\,(\ref{eq-staircase}). Note that $\tilde{I}_{s}(0) =-2/\pi \approx -0.637$ according to Eq.\,(\ref{eq-j-beta0}).}
\end{figure}
and shown in Fig.\.5. Let us remind the reader that the positive and negative signs of $j$ and $I$ mean the counterclockwise and clockwise currents, respectively. As follows from Eq.\,(\ref{eq-j-beta0}), the total current in a disk ($\beta =0$) is $\tilde{I}_{s}(0)=-2/\pi $. 

Now the perforating fields $h_{n}$ at $n \geq 2$, see Fig.\,3, can be found using a simple recurrent algorithm. The case $n=1$ requires a separate consideration and will be discussed in Section IV. At the zeroth level ($h_{ext}<h_1$) of a staircase shown in Fig.\,3, the shielding current in a ring is 
\begin{equation*}
\tilde{j}_{0}(h_{ext})=h_{ext}(\tilde{j}_{s}-\tilde{j}_{p}).
\end{equation*}%
After the first perforation occurs at certain $h_{ext}=h_{1}$, the current $\tilde{j}(\zeta)$ is changed to fit the condition $\tilde{I}=0$. Thus, when pumping of magnetic flux inside the ring by the first perforating avalanche is over, we get:%
\begin{equation}
\tilde{j}_{1}(h_{1}) = h_{1}\left( \tilde{j}_{s}-\frac{\tilde{I}_{s}}{\tilde{I}_{p}}\tilde{j}_{p}\right),\quad
\tilde{\Phi}_{1} = h_{1}\tilde{\Phi}_{p}\frac{\tilde{I}_{p}-\tilde{I}_{s}}{\tilde{I}_{p}}.
\label{eq-j1}
\end{equation}%
where $\tilde{\Phi}_{1}$ is the penetrated flux. It should be emphasized that the mean field inside the ring, $\left\langle h\right\rangle =\tilde{\Phi}_{1}/\tilde{\Phi}_{p}$, is \emph{greater} than $h_{ext}$. This is just an illustration of a flux focusing effect discussed in Refs.\,\cite{BrandtClem,Brojeny}. Its explanation is very simple: the total current $\tilde{I}=0$ is a sum of a negative $\tilde{I}_{-}$ flowing at the outer part of a ring and positive $\tilde{I}_{+}$ in its inner area, similar to the current distribution shown in Fig.\,1. These two currents are equal by an absolute value, $\left\vert \tilde{I}_{-}\right\vert =\left\vert \tilde{I}_{+}\right\vert $, but $\tilde{I}_{+}$ is closer to the ring center. Thus the total flux produced by the shielding currents is positive, resulting in an enhancement of the mean field inside the ring: $\left\langle h\right\rangle >h_{ext}$. In a cylindrical geometry the flux focusing effect is absent in the $I=0$ state, and $\left\langle h\right\rangle =h_{ext}$, but in the case of a flat ring the focusing effect is quite pronounced \cite{BrandtClem,Brojeny}, especially in a thick ring ($\beta \ll 1$), where $(\tilde{I}_{p}-\tilde{I}_{s})/\tilde{I}_{p}) \rightarrow \infty$, see Eq.\,(\ref{eq-j1}) and Fig.\,5. 

Until the next perforation, while we remain at the first level $h_1<h_{ext}<h_2$ of the staircase shown in Fig.\,3, vortices do not penetrate inside a ring and $\tilde{\Phi}$, as well as $\left\langle h\right\rangle$, remain unchanged. Therefore, we need to find such a dependence of $\tilde{j}$ on $h_{ext}$, which obeys the condition $\tilde{\Phi}(h_{ext})=\tilde{\Phi}_{1}=\mathrm{const}$.
The solution, which meets this condition and provides, of course, $h=0$ inside the ring body $\beta<\zeta<1$, is%
\begin{equation*}
\tilde{j}_{1}(h_{ext})=h_{ext}(\tilde{j}_{s}-\tilde{j}_{p})+h_{1}\frac{%
\tilde{I}_{p}-\tilde{I}_{s}}{\tilde{I}_{p}}\tilde{j}_{p}.
\end{equation*}%
Correspondingly, 
\begin{equation}
\tilde{I}_{1}(h_{ext})=\left( h_{ext}-h_{1}\right) (\tilde{I}_{s}-\tilde{I}%
_{p})<0  \label{eq-I1}
\end{equation}%
within the first level of a staircase. The next perforation (the jump from the first level to the second one) occurs, according to Eq.\,(\ref{eq-perfor}), at $\tilde{I}_{1}(h_{ext})=-q$. As follows from Eq.\,(\ref{eq-I1}), it happens at 
\begin{equation*}
h_{ext}=h_{2}=h_{1}+\frac{q}{\tilde{I}_{p}-\tilde{I}_{s}},
\end{equation*}%
see Fig.\,3. Immediately after this (second) perforation takes place, an $\tilde{I}=0$ state is formed again, which means%
\begin{equation*}
\begin{split}
\tilde{j}_{2}(h_{2}) =h_{2}\left( \tilde{j}_{s}-\frac{\tilde{I}_{s}}{%
\tilde{I}_{p}}\tilde{j}_{p}\right) , \\
\tilde{\Phi}_{2} =h_{2}\tilde{\Phi}_{p}\frac{\tilde{I}_{p}-\tilde{I}_{s}}{%
\tilde{I}_{p}}=\tilde{\Phi}_{1}+q\frac{\tilde{\Phi}_{p}}{\tilde{I}_{p}},
\end{split}
\end{equation*}%
and so on, as shown in Fig.\,3. 

The general solution of this recurrent problem can be written as%
\begin{equation}
\begin{split}
h_{n} = h_{1}+\frac{(n-1)q}{\tilde{I}_{p}-\tilde{I}_{s}};
 \\
\left\langle h\right\rangle _{n} =\frac{\tilde{\Phi}_{n}}{\pi \beta ^{2}}%
=h_{n}\frac{\tilde{I}_{p}-\tilde{I}_{s}}{\tilde{I}_{p}},
\end{split}
\label{eq-staircase}
\end{equation}%
for $n \geq 2$, see Fig.\,3. Here $h_n$ is the external field corresponding to the $n$-th perforation (starting from the second one), $\tilde{\Phi}_{n}$ and $\left\langle h\right\rangle _{n}$ are respectively the trapped flux and mean trapped field inside a ring after the $n$-th perforation is over ($n \geq 2$). The width of steps $\Delta h=h_{n+1}-h_{n}=q/\left( \tilde{I}_{p}-\tilde{I}_{s}\right) $ increases as a function of $\beta $, and ranges between $\pi/2=1.57$ at $\beta \rightarrow 0$ and $2.61$ at $\beta =0.95$, see Fig.\,6. It is problematic to carry out calculations at $\beta >0.95$ due to divergence of $j$ near both inner and outer edges of a ring, which approach each other at $\beta \rightarrow 1$. 
\begin{figure}[t!]
    \centering
    \includegraphics[width=0.48\textwidth]{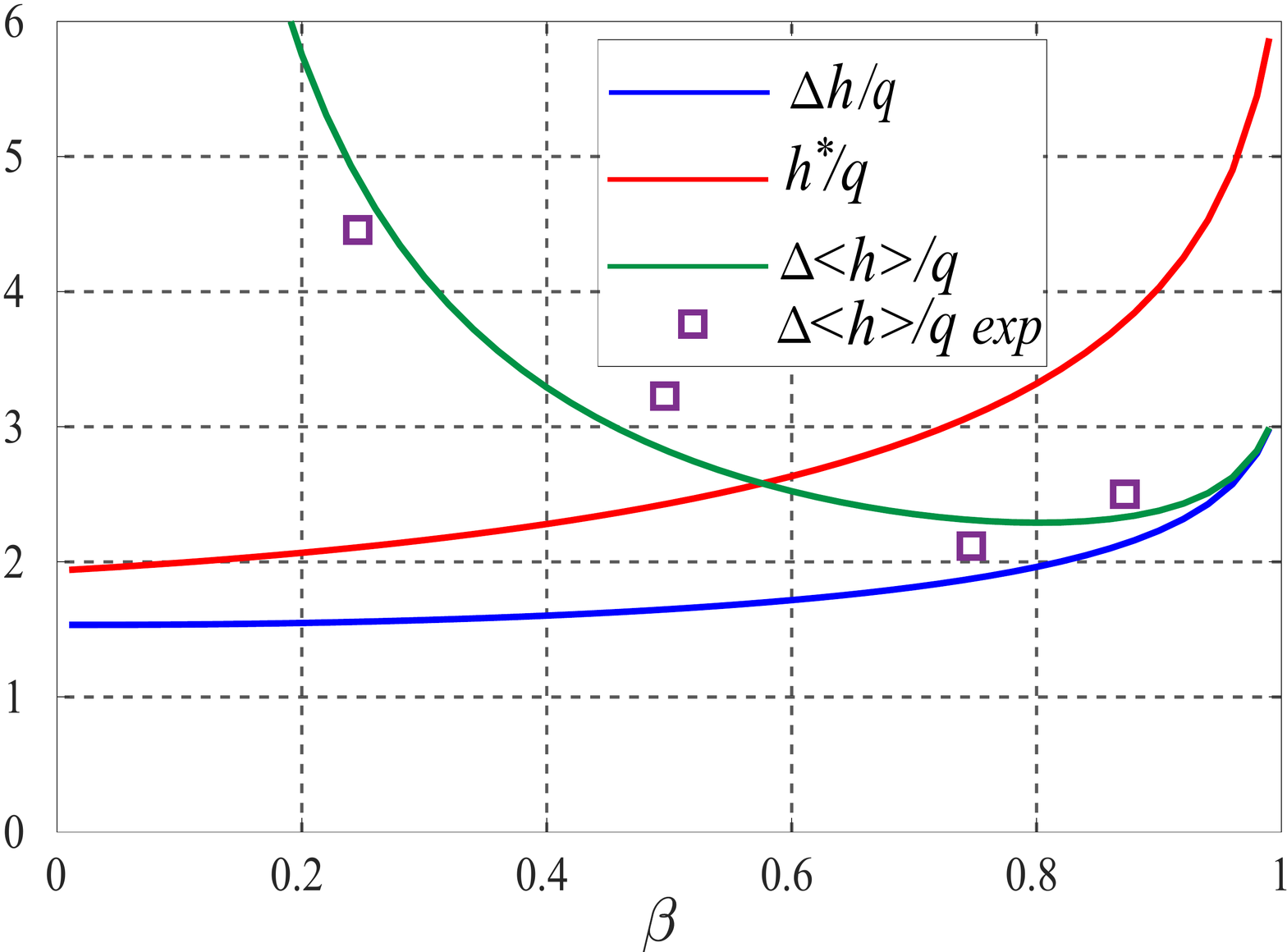}
    \caption{The width, $\Delta h/q$, and height, $\Delta \left\langle h\right\rangle/q$, of steps shown in Fig.\,3 (except the first one) as functions of $\beta$. All the values shown in this graph are dimensionless. The field $h^\ast$ serves as the lower boundary for $h_1$. Experimental data for $\Delta\left\langle H\right\rangle$ \cite{Shvartzberg} are shown in squares using $q=35$, see discussion in Section VII.}
\end{figure}
The step height, starting from the second one, is $\Delta \left\langle h\right\rangle =q/\tilde{I}_{p}$ according to Eq.\,(\ref{eq-staircase}). It is rather curious that $\Delta \left\langle h\right\rangle $ depends non-monotonously on $\beta $, see Fig.\,6.

Note that the dimensionless values for fields (and, correspondingly, steps of the staircase structure) used above are related to the original units just by the factor $H_0$ defined by Eq.\,(\ref{eq-H0j0}). We will discuss this relation in greater detail in Section VII, which is devoted to comparison of our theoretical results to the experimental data.

\section{The first perforating field $h_1$}

The field $h_1$ where the first flux perforation into the central area of a ring takes place, see Fig.\,3, remains undefined in the previous section and requires a separate analysis. It appears to be inevitably greater than the width $\Delta h$ of further steps (levels), see Fig.\,3. In fact, the condition set by Eq.\,(\ref{eq-dE0}) is satisfied at $h_{ext}=\Delta h$, but $E(\zeta )$ is a monotonously decreasing function with a maximum at $\zeta =\beta $, see the red line in Fig.\,7 and compare it with the green line in Fig.\,2, which obeys Eq.\,(\ref{eq-dE0}) as well.
\begin{figure}[t!]
    \centering
    \includegraphics[width=0.48\textwidth]{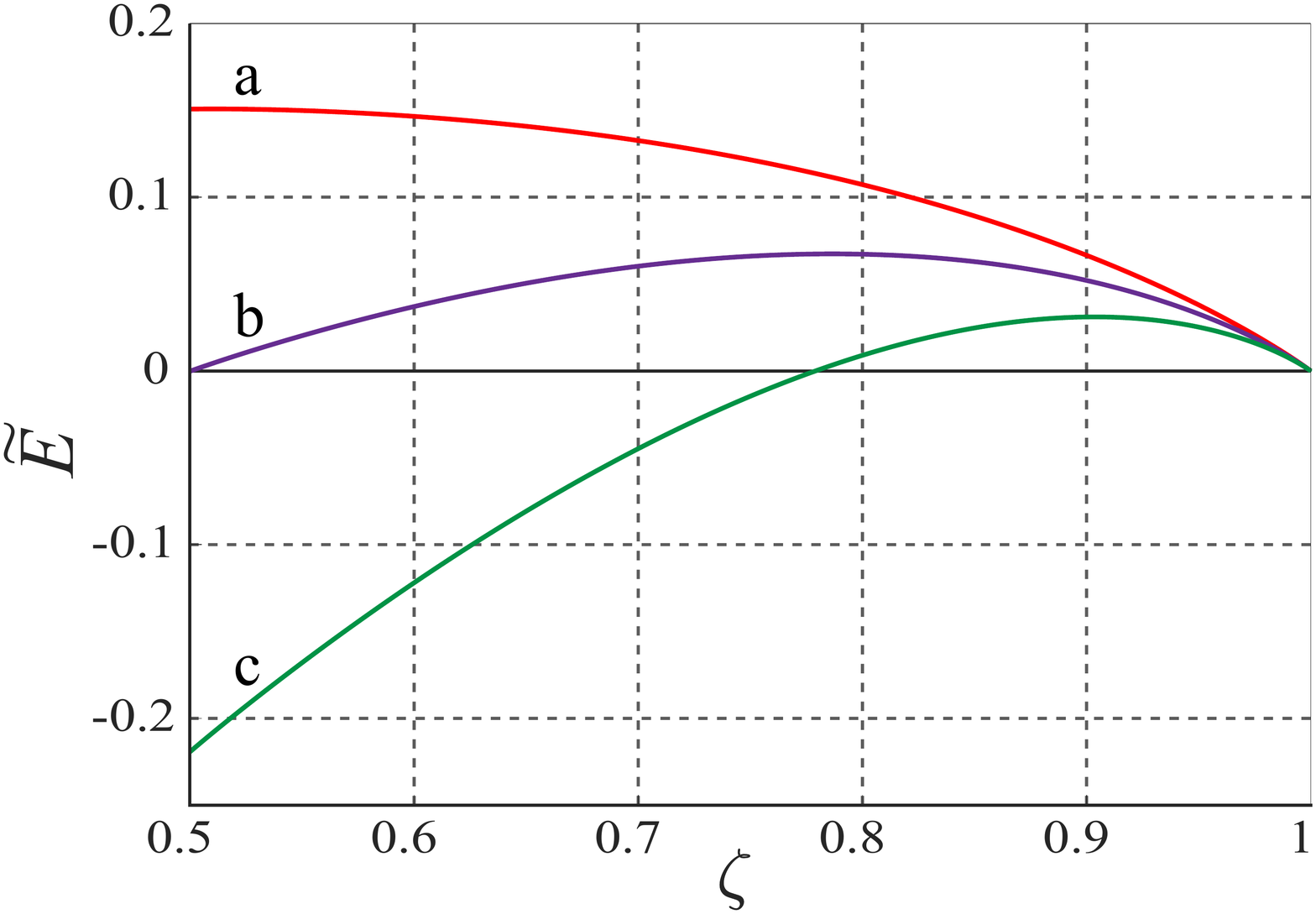}
    \caption{Energy $\tilde{E}(\zeta)$, see Eq.\,(\ref{eq-Etilde}), of the first penetrating finger for $\beta=0.5$ (full shielding, $h<h_1$). (a) The perforation condition, see Eq.\,(\ref{eq-perfor}), is formally fulfilled, $h_{ext}=\Delta h$, $d\tilde{E}/d\zeta \mid _{\zeta =\beta}=0$, but $\tilde{E}>0$ at all $\zeta$, no penetration; (b) $\tilde{E}(\beta)=0$, $h_{ext}= h^\ast$; (c) $h_{ext}=h_1> h^\ast$, potential barrier is surmountable, a first perforating avalanche appears. }
\end{figure}
Apparently, the field $h^{\ast}$ determined by a condition $E(\beta )=0$, see the violet line in Fig.\,7, sets the lower boundary for $h_1$. It is necessary to increase $h_{ext}$ above $h^{\ast }$ until the surface barrier becomes surmountable for a dendrite (finger) emerging from the outer edge, see the green line in Fig.\,7. So the width of the zeroth level of a staircase, $h_1$, is significantly greater than $\Delta h$ (see Fig.\,3). It is determined exclusively by the outer surface (edge) of a ring, especially by presence of defects and imperfections, which locally diminish the surface barrier and facilitate the entrance of vortices. A similar phenomenon was reported in bulky superconducting samples \cite{Burlachkov-defects}, where the effect of local breaking of the Bean-Livingston barrier was considered. The field $h^{\ast }(\beta )$ is shown in Fig.\,6 and serves as a lower boundary for $h_1(\beta)$, but the actual value of $h_{1}$ is expected to be sample-dependent in contrast to further perforating fields $h_{n\geq 2}$, which are determined by the ring geometry, see  Eq.\,(\ref{eq-staircase}). It can be checked numerically that the surface barrier at $h_{n\geq 2}$ is less than at $h_{1}$ and decreases with $n$, so if a barrier is surmountable for the first perforation, it will be all the more surmountable at all the next perforating fields determined by Eq.\,(\ref{eq-staircase}).

The general scenario of a staircase flux trapping by a pinningless superconducting ring looks as follows. At $h_{ext}=h_{1}>h^{\ast }$ a propagating finger overcomes a surface barrier, lightens inside the ring, reaches its inner edge and establishes a magnetic bridge across the ring. Note that the first penetration is always pass-through (edge to edge): once a finger has surmounted the surface barrier, it is pushed by a shielding current straight inside until it reaches the inner edge. This results from the fact that $\tilde{E}(\zeta)$ has no any intermediate minimum for a perfect Meissner state, see the green curve in Fig.\.7. The surface barrier for the next penetrations $h_{n \geq 2}$, is significantly less than that at $h_{1}$. Therefore, the next perforating finger should not necessarily originate from the same point at the outer edge where the first avalanche started. Also, at $n\geq 2$, a partial dendrite penetration is possible, where an avalanche starts to propagate but stops in an intermediate position $\zeta_0$ inside the ring body, see the blue curve in Fig.\,2. As $h_{ext}$ grows up further, $\zeta_0$ moves to the left in Fig.\,2 and finally reaches $\beta $. Then the function $E(\zeta)$ becomes monotonous, see the green line in Fig.\,2, and the next edge to edge perforation takes place.

As follows from Eq.\,(\ref{eq-staircase}), a staircase function $\left\langle h\right\rangle $ vs. $h_{ext}$ is confined between the $\tilde{I}=0$ line%
\begin{equation}
\left\langle h\right\rangle _{\tilde{I}=0}=\frac{\tilde{I}_{p}-\tilde{I}_{s}%
}{\tilde{I}_{p}}h_{ext}  \label{eq-hI0}
\end{equation}
at the top, and the "perforation line"%
\begin{equation}
\left\langle h\right\rangle _{perf}=\frac{\tilde{I}_{p}-\tilde{I}_{s}%
}{\tilde{I}_{p}}h_{ext}-\frac{q}{\tilde{I}_{p}}  \label{eq-hperf}
\end{equation}
at the bottom, see Fig.\,3. The slope of both lines $\left( \tilde{I}_{p}-\tilde{I}_{s}\right) /\tilde{I}_{p}$ is greater than $1$: it approaches $1$ at $\beta \rightarrow 1$ (linear circle) and diverges at $\beta \rightarrow 0$ (solid disk). 

\section{Energy dissipated by a perforating dendrite}

Using a recurrent procedure described in Section III, at the moment of the $n$-th perforation, when the condition set by Eq.\,(\ref{eq-perfor}) is fulfilled, $\tilde{j}$ can be expressed as%
\begin{equation}
\tilde{j}(h_{n})=h_{n}(\tilde{j}_{s}-\tilde{j}_{p})+\frac{h_{1}\left( \tilde{%
I}_{p}-\tilde{I}_{s}\right) +(n-1)q}{\tilde{I}_{p}}\,\tilde{j}_{p}.
\label{eq-jhn}
\end{equation}
Let us denote%
\begin{equation}
\begin{split}
\tilde{E}_{p}(\beta ) =\int_{\beta }^{1}d\zeta \int_{\zeta }^{1}d\zeta
_{1}\,\,\tilde{j}_{p}(\zeta _{1}),  \\
\tilde{E}_{s}(\beta ) =\int_{\beta }^{1}d\zeta \int_{\zeta }^{1}d\zeta
_{1}\,\,\tilde{j}_{s}(\zeta _{1}).
\end{split}
 \label{eq-Eps}
\end{equation}%
Obviously, $\tilde{E}_{p}(\beta )>0$ and $\tilde{E}_{s}(\beta )<0$ for all $\beta $. These functions are plotted in Fig.\,8.
\begin{figure}[t!]
    \centering
    \includegraphics[width=0.48\textwidth]{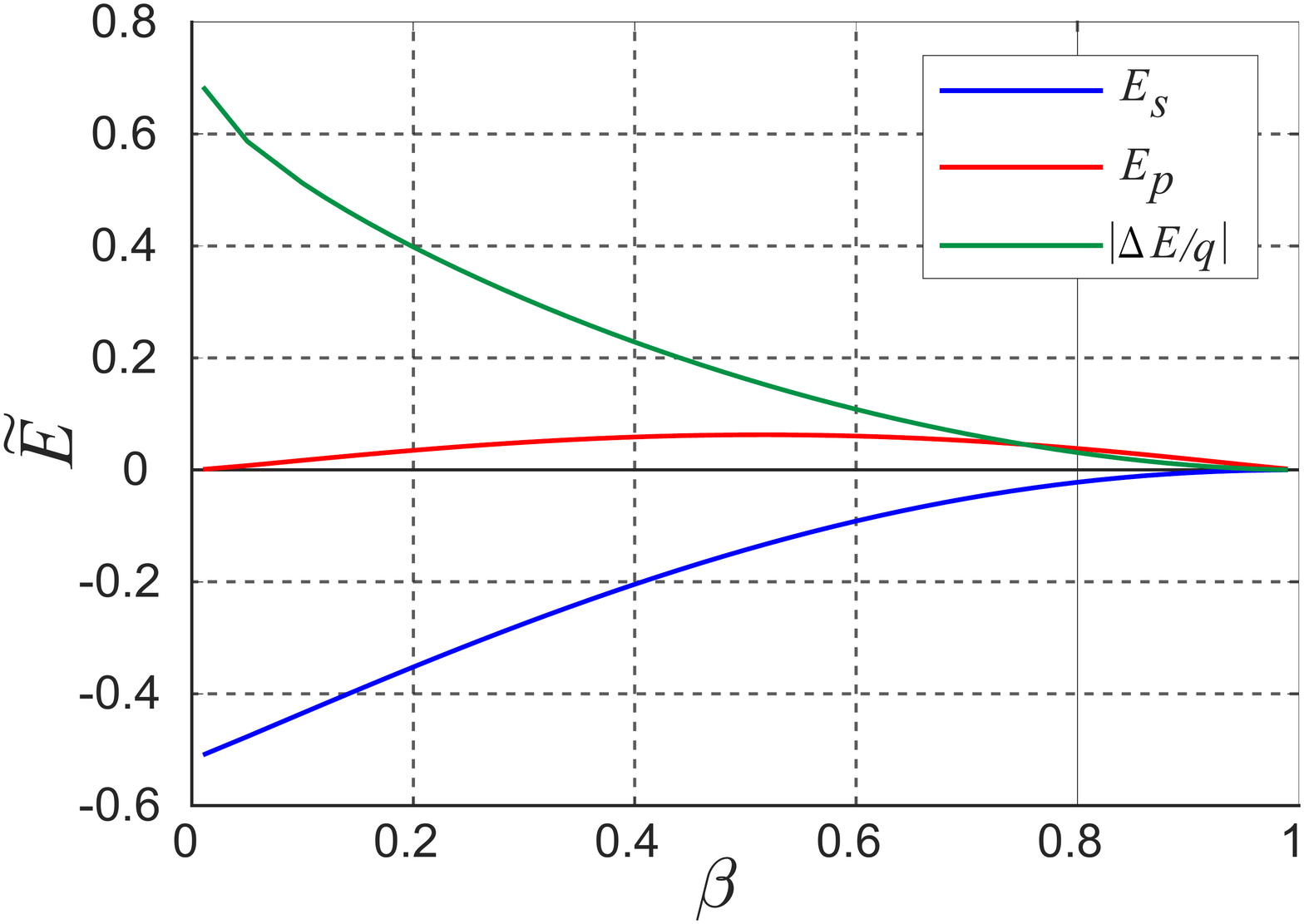}
    \caption{The energies $\tilde{E}_{p}(\beta )$ and $\tilde{E}_{s}(\beta )$ as functions of $\beta$. The dissipated heat grows by $\left\vert \Delta \tilde{E}\right\vert$ with each next avalanche.}
\end{figure}
Substituting Eqs.\,(\ref{eq-jhn}) and (\ref{eq-Eps}) into Eq.\,(\ref{eq-Etilde}), we find the (negative) energy of the $n$-th perforating avalanche:%
\begin{equation}
\tilde{E}_{n} = q(1-\beta )+h_n(\tilde{E}_s-\tilde{E}_p)+
\frac{h_1\left( \tilde{I}_{p}-\tilde{I}_{s}\right) +(n-1)q}{\tilde{I}_{p}}\,\tilde{E}_{p}
\label{eq-En}
\end{equation}
and the difference%
\begin{equation}
\Delta \tilde{E}=\tilde{E}_{n+1}-\tilde{E}_{n}=\frac{q}{\tilde{I}_{p}-\tilde{%
I}_{s}}\left( \tilde{E}_{s}-\frac{\tilde{I}_{s}}{\tilde{I}_{p}}\tilde{E}%
_{p}\right) .  \label{eq-deltaE}
\end{equation}%
The heat dissipated at the $n$-th penetration is $\left\vert \tilde{E}_{n}\right\vert $. As $n$ grows up, the released heat (a thermomagnetic effect which accompanied each avalanche) increases linearly. The factor $\left\vert \Delta \tilde{E}/q\right\vert$ is shown in Fig.\,8 as a function of $\beta $. It is quite natural that the thermomagnetic effect grows with the ring thickness, which is proportional to $1-\beta$. Note that the dimensionless energies determined in this Section are related to real units ($\mathrm{erg}$) by a factor of $E_0$, see Eq.\,(\ref{eq-E}).

\section{Effect of pinning - unsaturated perforation}

Above we considered a "clean" pinningless ring and got a staircase picture of flux penetration shown in Fig.\,3 and described by Eq.\,(\ref{eq-staircase}). Now let us discuss the effect of pinning. If there exists a pinning force $f$ per unit length of a moving vortex, then the total "friction" force $F$ which acts on a propagating dendrite of length $a-r$, see Fig.\,1, is $F=f \rho ws(a-r)$. Note that $f$ is related to the critical current as $j_c=cf/\phi_0$. It is worth emphasizing that $j_c$ is a "true" critical current density, measured in $\mathrm{A/cm^2}$, whereas the current density $j$ defined above in our paper is the density of a linear current and measured in $\mathrm{A/cm}$. Then the condition of perforation is $dE/dr|_{r=b}=F$, which can be rewritten in the dimensionless coordinates as
\begin{equation}
\tilde{I}_{perf}=\int_{\beta }^{1}\,\,\tilde{j}(\zeta )\,d\zeta =-q-\tilde{f},
\label{eq-perfor-pin}
\end{equation}%
where $\tilde{f}=4\pi f(a-b)/\phi_0 H_{c1}$. Comparing Eq.\,(\ref{eq-perfor-pin}) with Eq.\,(\ref{eq-perfor}), we see that in what concerns the perforation condition, pinning just effectively increases the parameter $q$. It means that the perforation line, see Fig.\,3, is located farther from the $I=0$ line. The first step should be also increased by pinning, but its size is determined by the activation over the surface barrier, and a quantitative description of such an increase is beyond the scope of our paper. However, the width and height of steps in the staircase picture, see Fig.\,3, are affected dramatically in presence of pinning. As a superconducting bridge is established, vortices will penetrate along it to the inner part of a ring until the positive work of superconducting currents $j$ on a vortex who travels throughout the whole finger from $r=a$ to $r=b$ remains greater than the absolute value of the negative work produced by pinning force on the same vortex. Thus, penetration will end up when the total current $I=I_{pin}$, where
\begin{equation}
I_{pin} =-f\frac{(a-b)c}{\phi_0}=-j_c (a-b)s.
\label{eq-I-pin}
\end{equation}
We wrote Eq.\,(\ref{eq-I-pin}) in real (not dimensionless) valuables since this way it becomes most clear. As a result, the steps in the $\left\langle h\right\rangle $ vs. $h_{ext}$ dependence become smaller and no longer "touch" the $I=0$ line, see Fig.\,9.
\begin{figure}[t!]
    \centering
    \includegraphics[width=0.48\textwidth]{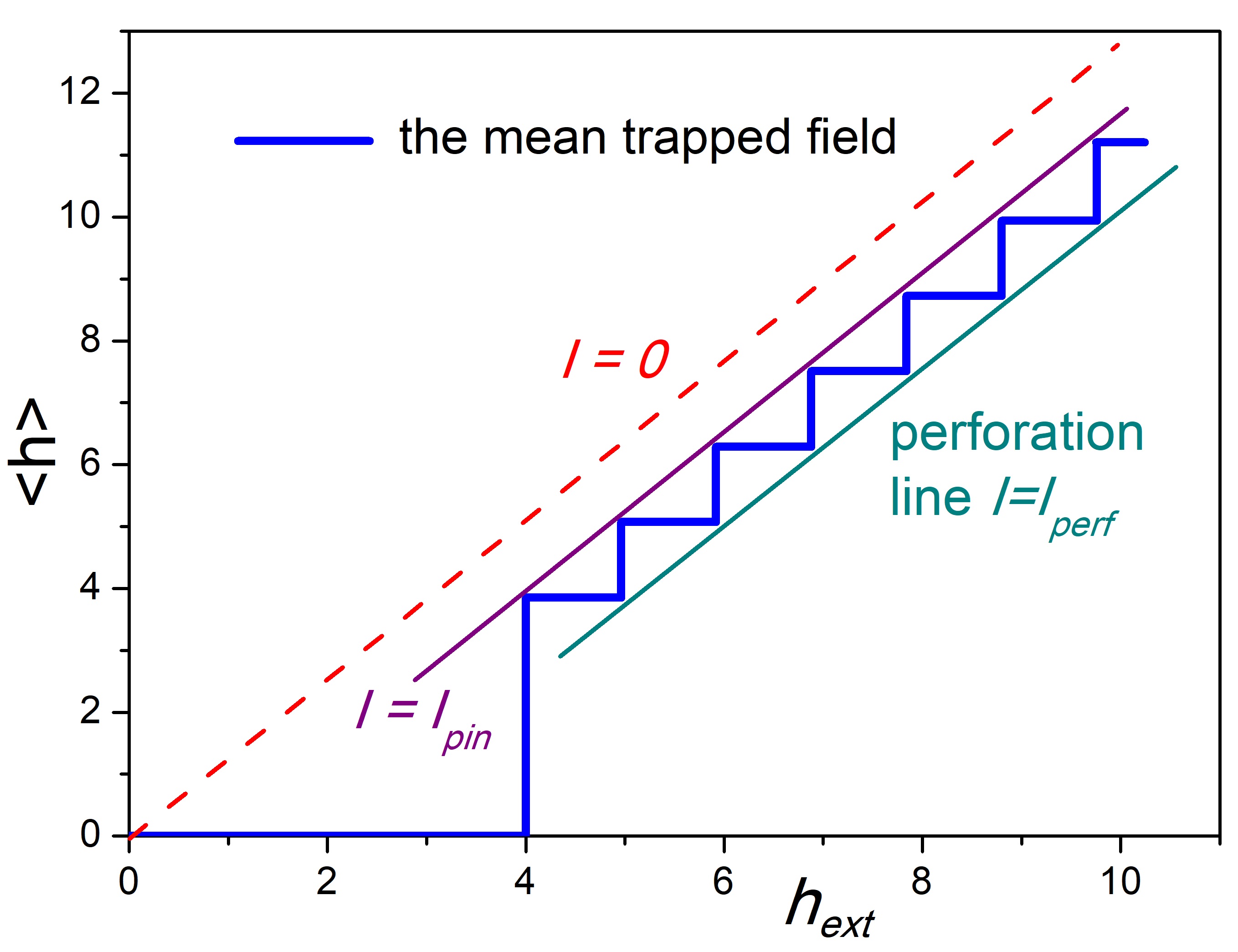}
    \caption{The mean field trapped in the ring $\left\langle h\right\rangle $ as a function of $h_{ext}$ at $\beta=0.75$ in a sample with pinning - unsaturated penetration. The steps do not "grow up" till $I=0$ line and stop at $I=I_{pin}$, compare to Fig.\,3.}
\end{figure}
Such a staircase, where pinning prevents the bridge (perforating avalanche) to "saturate" the magnetic field inside the ring (like happens in the pinningless case considered in Section III) and reach the zero current state $I=0$, can be called non-saturated.

\section{Comparison with the experimental data}

There are only a few experimental studies of dendrite penetration into a thin superconducting ring \cite{Olsen1,Shvartzberg,Jiang}, whereas the papers on magnetic avalanches in thin disks, squares and strips are numerous. Most of our theoretical results are in a good concordance with the experimental data. A staircase structure where the first step is significantly greater than the successive ones was observed in all the experiments mentioned above. In Ref.\,\cite{Shvartzberg} a set of four identically fabricated $\mathrm{Nb}$ rings of thickness $s=200\,\mathrm{nm}$ was used. The outer radii of all rings was $a=400\,\mathrm{\mu m}$, whereas the inner radii $b$ were $50$, $100$, $200$ and $300$\thinspace $\mathrm{\mu m}$, which corresponds to $\beta =1/8,\,1/4,\,1/2$ and $3/4$, respectively. A staircase dependence $\left\langle h\right\rangle $ vs. $h_{ext}$ presented in Ref.\,\cite{Shvartzberg}, where all the steps touch the $I=0$ line, suggests that the $\mathrm{Nb}$ samples are clean enough and pinning is negligible. The width of steps (starting from $n=2$) can be estimated as $\Delta H=H_{0}\,\Delta h\approx $ $25\,\mathrm{Oe}$ at $\beta =1/2$ and $T=5\,\mathrm{K}$ \cite{Shvartzberg}. Since $H_{0}=sH_{c1}/2a\approx 0.4\, \mathrm{Oe}$ and $\tilde{I}_{p}-\tilde{I}_{s}\approx 0.6$ at $\beta =1/2$, see Fig.\,5, using Eq.\,(\ref{eq-staircase}) we get $q \approx 35$. As was mentioned above,  $q\varepsilon_{0}$ is the  energy of a straight finger per one vortex. Then, according to a classic work on the energy of an Abrikosov vortex lattice \cite{Fetter}, in $\mathrm{Nb}$-like superconductors with $\kappa \simeq 1$ we can estimate $q \gtrsim 1+2\pi (\lambda /d)^{2}$, where $d$ is the mean intervortex distance and the sign $\gtrsim$ accounts for bending and branching of a propagating avalanche. For dendrites observed in Ref.\,\cite{Shvartzberg}, we find $\lambda /d \lesssim 2.3$, which looks quite reasonable. 

As was mentioned in Section III, the mean height of steps $\Delta \left\langle h\right\rangle =q/\tilde{I}_{p}$ depends non-monotonously on $\beta $ with a minimum at $\beta \approx 0.83$, see Fig.\,6. Such a dependence was observed experimentally: the values for $\Delta \left\langle h\right\rangle (\beta )$ reported in Ref.\,\cite{Shvartzberg} are compared in Fig.\,6 with our theoretical results using $q = 35$ as estimated above. The accordance is good, especially taking into account that the experimental $\Delta \left\langle h\right\rangle $ was calculated over all the steps including the first step which, as follows from our analysis, should be excluded from the statistics for $\Delta \left\langle h\right\rangle $.

When discussing the height of experimentally measured staircase steps, it is worth emphasizing that each perforating avalanche consists of thousands of vortices at least. For instance, $\Delta \left\langle H\right\rangle \approx 20 \,\mathrm{Oe}$ in a ring with the inner radius $b=200\,\mathrm{\mu m}$, see Fig.\,3 in Ref.\,\cite{Shvartzberg}, corresponds to a simultaneous penetration of at least $10^4$ flux quanta.

An example of a non-saturated flux penetration described in Section VI is presented in Ref.\,\cite{Olsen1}, where flux penetration in a $\mathrm{MgB}_{2}$ ring with $a=2\mathrm{\,mm}$ and $b=1.1\mathrm{\,mm}$ ($\beta =0.55$) was studied at various temperatures. The slope of all the staircases $\left\langle h\right\rangle $ vs. $h_{ext}$ extracted for the experimental data (see Fig.\,2 in Ref.\,\cite{Olsen1}) was approximately $1.41$, whereas our prediction is $\left( \tilde{I}_{p}-\tilde{I}_{s}\right) /\tilde{I}_{p}\approx 1.54$. This can be considered as a reasonably good fit. As temperature increases, pinning is fading out and the width and height of steps grow up in a complete concordance with our theory. Note also that the magneto-optical images of dendrite formation presented in experimental studies \cite{Shvartzberg} ($\mathrm{Nb}$) and \cite{Olsen1} ($\mathrm{MgB}_{2}$) are qualitatively different. In $\mathrm{Nb}$ at any stage only a few partially penetrated (with $r>b$, see Fig.\,1) dendrites were observed, thus most of the ring body was free of magnetic flux at any stage of experiment. In $\mathrm{MgB}_{2}$ all the outer half of the ring is filled up with numerous partially penetrated dendrites adjacent to each other. This confirms that pinning in $\mathrm{Nb}$ is negligible, whereas in $\mathrm{MgB}_{2}$ it is significant, at least at low temperatures. 

\section{Conclusion}

We built up a quantitative description of a magnetic avalanche penetration into a superconducting ring. Using a recurrent calculation procedure, a staircase dependence of the mean field trapped inside a ring on the external field is explained, where the first step is considerably greater than the following ones in complete accordance with available experimental data. The width of the first step (the zeroth level of a staircase) is determined by a surface barrier for avalanche emergence at the outer edge of a ring, whereas the geometry of the successive steps (starting from the second one), as well as the staircase slope, depend on the ring geometry. The heat released at each next perforation grows linearly with the number of step. Flux pinning, if present, results in an unsaturated magnetic penetration observed experimentally in $\mathrm{MgB}_{2}$ samples, where steps become smaller than in a pinningless case. As temperature increases, the effect of pinning fades out, but the slope of a staircase remains the same since it is determined only by the ring shape (ratio of the inner and outer radii). All our results are expressed as functions of this ratio and are applicable to flat rings of any geometry. Our theory should be useful for description of available experimental data (so far its accordance with available experimental data is quite promising), planning new experiments and, hopefully, for design of digitizing devices based on a step-like flux penetration into superconducting rings.

\section{Acknowledgments}
We thankfully acknowledge useful conversations with Y.\,Yeshurun, A.\,Shaulov and Y.\,Shvartzberg.

\end{document}